\documentclass[aip,apl,twocolumn,amsmath,amssymb,bibnotes]{revtex4}

\usepackage{float}
\usepackage{color,epsfig,rotating,graphicx,subfigure}

\usepackage{amsmath,amssymb,amscd,ulem,comment}

\usepackage[latin1]{inputenc}
\usepackage[english]{babel}
\usepackage{tikz}
\usepackage{dsfont}
\usepackage{placeins}

\renewcommand{\Re}{{\rm Re}}
\renewcommand{\Im}{{\rm Im}}

\newcommand{\ri}{{\rm i}}

\newcommand{\kb}{k_{\rm B}}

\newcommand{\alphamat}{\underline{\underline{\alpha}}}
\newcommand{\chimat}{\underline{\underline{\chi}}}
\newcommand{\blockt}{\boldsymbol{T}^{-1}}
\newcommand{\I}{{\rm i}}
\newcommand{\mue}{{\rm \mu}}
\begin{document}
\title{Thermal rectification and spin-spin coupling of non-reciprocal localized and surface modes}
\date{\today}

\author{Annika Ott and Svend-Age Biehs$^{*}$}
\affiliation{Institut f\"{u}r Physik, Carl von Ossietzky Universit\"{a}t, D-26111 Oldenburg, Germany}
\email{s.age.biehs@uni-oldenburg.de} 

\begin{abstract} 
We study the rectification of near-field radiative heat transfer between two InSb nano-particles due to the presence of non-reciprocal surface modes in a nearby InSb sample when an external magnetic field is applied and its dependence on the magnetic field strength. We reveal the spin-spin coupling mechanism of the localized particle resonances and the surface mode resonances which is substantiated by the directional heat flux in the given setup.  We discuss further the interplay of the frequency shift, the propagation length, and local density of states on the strength and directionality of the rectification as well as the non-reciprocal heating effect of the nanoparticles. 
\end{abstract}


\maketitle

\section{Introduction}

In the last few years, it could be shown that the non-reciprocal behaviour of magneto-optical materials like InSb has very interesting consequences for nanoscale thermal radiation. For example, fundamental effects like a persistent heat-current~\cite{zhufan,zhufan2}, giant magneto-resistance~\cite{Latella2017,Cuevas}, thermal Hall effect~\cite{hall} as well as a circular heat flux, angular momentum, and spin which do also persist in global equilibrium~\cite{meinpaper} were highlighted. As reviewed and discussed in detail in Ref.~\cite{OttEtAl2019} magneto-optical materials and in particular these fundamental effects might have applications in the control of magnitude~\cite{moncadavilla,Song,WuEtAl} and direction of radiative heat fluxes in nanoscale systems. Furthermore, it could be shown that non-reciprocal materials can also be utilized to introduce a near-field heat flux rectification. So far, most of the concepts for thermal rectification are based on the temperature dependence of the material properties~\cite{FanRectification2010,Iizuka2012,BasuEtAl2011,WangEtAl2013,Nefzaoui2014,OrdonezEtAl2017} which can be very strong for phase-change materials like VO$_2$~\cite{QazilbashEtAl2007} which show up to date the strongest diode effect for thermal radiation~\cite{PBASAB2013,Yangetal2013,ItoEtAl,FiorinoEtAl2018}. Recently, it has been demonstrated that a thermal emitter and receiver can show an enhanced heat exchange by transporting the heat via the surface modes of a third body in their close vicinity~\cite{Saaskilathi2014,Asheichyk2017,DongEtAl2018,paper_2sic,ZhangEtAl2019,HeEtAl2019,MessinaEtAl2012,MessinaEtAl2016,ZhangEtAl2019b} or by coupling to large wave-vector propagating modes in hyperbolic materials which is very similar to the F\"{o}rster resonance energy transfer enhancement observed in plasmonic and hyperolic environments~\cite{Foerster,BiehsEtAl2016,DeshmukhEtAl2018,NewmanEtAl2018}. This coupling effect opens up a new possibility to rectify the radiative heat flux between a thermal emitter and receiver by introducing surfaces supporting non-reciprocal surface modes~\cite{BrionEtAl1972,wallis,chiuquinn}. As has been shown by us, non-reciprocal surface modes allow to rectifiy the radiative heat flux very efficiently~\cite{paper_diode}.  

\begin{figure}
	\parbox{0.5\textwidth}{
		\centering	
		\begin{tikzpicture}[transform shape,scale=1.2]
		\filldraw [fill=blue ,draw=black] (2,0)  rectangle  (7.7,1);
		\node[above=0 mm, right=0mm] at (3.6,0.5,0.0) {\textcolor{white}{\scalebox{1.5}{\textbf{Substrate}}}};
		\filldraw [above=17 mm, right=18mm, fill=red ,draw=black] (2,0) circle (5 mm);
		\filldraw [above=17 mm, right=39mm, fill=blue ,draw=black] (2,0) circle (5 mm);
		\draw[->, line width=5pt, draw=red ] (4.4,1.7) -- (5.3, 1.7);
		\node at (2.7,3.0,0.0) {\textcolor{black}{\scalebox{2}{\boldmath$\otimes$}}};
		\node[above=0.5mm, right=1.4 mm] at (2.7,3.0,0.0) {\textcolor{black}{\scalebox{1.3}{$\vec{B}$}}};
		\node at (4.7,2.4,0.0) {\textcolor{black}{\scalebox{1.3}{$P_2$}}};
		\node[above=17 mm, right=36mm] at (2.0,0.0,0.0) {\textcolor{white}{\scalebox{1.5}{$\boldmath{T_2}$}}};
		\node[above=17 mm, right=15mm] at (2.0,0.0,0.0) {\textcolor{white}{\scalebox{1.5}{$\boldmath{T_1}$}}};
		\node[above=0.0mm, right=0.0 mm] at (7,0.7,0.0) {\textcolor{white}{\scalebox{1.5}{$\boldmath{T_b}$}}};
		\node[above=0.0mm, right=0.0 mm] at (7,1.3,0.0) {\textcolor{black}{\scalebox{1.5}{$\boldmath{T_b}$}}};
		\node[above=0.0mm, right=0.0 mm] at (6.,3.0,0.0) {\textcolor{black}{\scalebox{1.2}{\textbf{forward}}}};
		\draw[->,line width=0.6mm,orange] (2.0,1.) --(3.0,1.);
		\draw[->,line width=0.6mm,orange] (2.2,0.8)--(2.2,1.8,0.0);
		\node[above=-0.5mm, right=-5.2 mm] at (2.2,1.8,0.0) {\textcolor{orange}{\scalebox{1.7}{$\mathbf{z}$}}};
		\node[above=-2.8mm, right=-2.5 mm] at (3.0,1.0,0.0) {\textcolor{orange}{\scalebox{1.7}{$\mathbf{x}$}}};

		\end{tikzpicture}
	\vspace{0.6cm}	
	}

	\parbox{0.5\textwidth}{
		\centering	
		\begin{tikzpicture}[transform shape,scale=1.2]
		\filldraw [fill=blue ,draw=black] (2,0)  rectangle  (7.7,1);
		\node[above=0 mm, right=0mm] at (3.6,0.5,0.0) {\textcolor{white}{\scalebox{1.5}{\textbf{Substrate}}}};
		\filldraw [above=17 mm, right=18mm, fill=blue ,draw=black] (2,0) circle (5 mm);
		\filldraw [above=17 mm, right=39mm, fill=red ,draw=black] (2,0) circle (5 mm);
		\draw[<-, line width=5pt, draw=red ] (4.4,1.7) -- (5.3, 1.7);
		\node at (2.7,3.0,0.0) {\textcolor{black}{\scalebox{2}{\boldmath$\otimes$}}};
		\node[above=0.5mm, right=1.4 mm] at (2.7,3.0,0.0) {\textcolor{black}{\scalebox{1.3}{$\vec{B}$}}};
		\node at (4.7,2.4,0.0) {\textcolor{black}{\scalebox{1.3}{$P_1$}}};
		\node[above=17 mm, right=36mm] at (2.0,0.0,0.0) {\textcolor{white}{\scalebox{1.5}{$\boldmath{T_2}$}}};
		\node[above=17 mm, right=15mm] at (2.0,0.0,0.0) {\textcolor{white}{\scalebox{1.5}{$\boldmath{T_1}$}}};
		\node[above=0.0mm, right=0.0 mm] at (7,0.7,0.0) {\textcolor{white}{\scalebox{1.5}{$\boldmath{T_b}$}}};
		\node[above=0.0mm, right=0.0 mm] at (7,1.3,0.0) {\textcolor{black}{\scalebox{1.5}{$\boldmath{T_b}$}}};
	    \node[above=0.0mm, right=0.0 mm] at (5.7,3.0,0.0) {\textcolor{black}{\scalebox{1.2}{\textbf{backward}}}};
				
		\draw[->,line width=0.6mm,orange] (2.0,1.) --(3.0,1.);
		\draw[->, line width=0.6mm,orange] (2.2,0.8)--(2.2,1.8,0.0);
		\node[above=-0.5mm, right=-5.2 mm] at (2.2,1.8,0.0) {\textcolor{orange}{\scalebox{1.7}{$\mathbf{z}$}}};
		\node[above=-2.8mm, right=-2.5 mm] at (3.0,1.0,0.0) {\textcolor{orange}{\scalebox{1.7}{$\mathbf{x}$}}};
		\end{tikzpicture}
	}
	\caption{Sketch of the considered system. Two InSb particles are in vicinity of an InSb substrate at a distance $z$. Top: Particle $1$ is heated with respect to particle $2$ and the background, i.e. the temperature $T_1 > T_2 = T_b$. Bottom: Reversed situation. Particle $2$ is heated with respect to particle $1$ and the background $T_2 > T_1 = T_b$.} 
	\label{Fig:Geometry}
\end{figure}
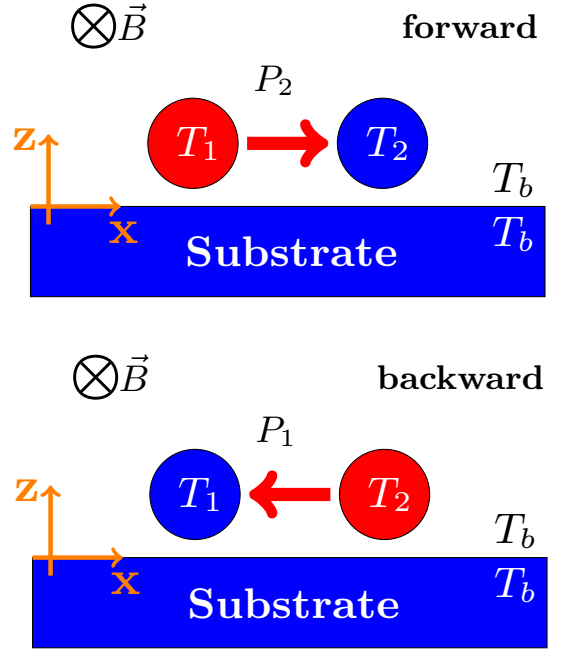

In this work, we will discuss our diode concept~\cite{paper_diode} as depicted in Fig.~\ref{Fig:Geometry} in much greater detail. In this configuration two InSb nanoparticles are held in close vicinity to an InSb substrate. The nano-particles can exchange heat via direct coupling or coupling to the surface modes. To quantify this heat exchange, we first derive the many-body expressions for the power exchanged between $N$ nano-particles in a given in general non-reciprocal environment as well as the many-body expression of the mean Poynting vector and discuss these quantities for the special case of $N=2$. We will show that the coupling to the surface modes is dictated by a spin-spin coupling mechanism which is behind the diode effect. We substantiate our interpretation by discussing the impact of the surface mode splitting, the propagation length, local density of states and the heat flux in the three-body structure as well as the thermal relaxation into the non-equilibrium steady state (NESS). We find, that the non-reciprocal heating of the nanoparticles can be on the order of 15\% of the applied temperature difference.

\section{heat flux}

In order to investigate heat flux rectification we consider the system shown in Fig.~\ref{Fig:Geometry}. Two identical spherical nanoparticles $1$ and $2$ with radius $R$ and interparticle distance $d$ are placed in plane parallel to a substrate at a distance $z$. We will assume that the particles are so small that we can describe them as dipoles with a polarizability $\uuline{\alpha}$. This assumption is valid for particles much smaller than the dominant thermal wavelength which is about $10\,\mu{\rm m}$ in our case and if the distances $d$ between the particles and $z$ between the particles and the substrate are at least $4R$~\cite{Otey,Becerril}. The particles and the substrate are made of a magneto-optical material supporting localized and surface resonances in the spectral window important for heat exchange at temperatures around $300\,{\rm K}$. In this work we chose InSb for the particles and for the substrate. Moreover, the background is vacuum. We assume throughout the work that the particles and their environment or background (substrate and vacuum background) can be assumed to be in local thermal equilibrium at temperatures $T_1$, $T_2$, and $T_b$. This assumption is valid as long as the radiative heat flux is less important than the heat conduction inside the materials, which is definitely fulfilled in our configuration. Furthermore, the assumption is only valid on time scales which are much larger than the time scale of thermal relaxation of the materials. This is again true since the heat conduction inside the materials is much larger than the heat conduction by thermal radiation between the particles and between the particles and the substrate.  

We can now determine the heat flux transferred between the two particles by calculating the net mean power $\langle P_i \rangle$ received by the $i$-th particle. We start here with the more general case that there are $N$ identical particles over a substrate following the derivations given in Ref.~\cite{nteilchen} and set $N = 2$ at the end. The total electric field $\vec{E}_i$ at the position of the $i$-th particle is given by the field contributions due to the fluctuating dipole moments $\vec{p}_j^{\,\,\rm fl}$ of all particles $j \neq i$ and the background field $\vec{E}_i^{\rm b}$ including direct thermal emission and multiple scattering. It can be written as~\cite{nteilchen}
\begin{equation}
    \begin{pmatrix} \vec{E}_{1} \\ \vdots \\ \vec{E}_{N} \end{pmatrix} =\boldsymbol{DT}^{-1}\begin{pmatrix} \vec{p}_1^{\,\,\rm fl} \\ \vdots \\ \vec{p}_N^{\,\,\rm fl}\end{pmatrix}+ (\boldsymbol{1}+\boldsymbol{DT}^{-1}\boldsymbol{A})\begin{pmatrix} \vec{E}_1^{\rm b} \\ \vdots \\ \vec{E}_N^{\rm b}\end{pmatrix}.
	    \label{Eq:electricfield}
\end{equation}
Similarly the induced dipole moments $\vec{p}_i$ for each particle $i$ can be expressed in term of the fluctuating dipole moments of all other particles and the background field~\cite{nteilchen}
\begin{eqnarray}
  \begin{pmatrix} \vec{p}_{1} \\ \vdots \\ \vec{p}_{N} \end{pmatrix} =\boldsymbol{T}^{-1}\begin{pmatrix} \vec{p}_1^{\,\,\rm fl} \\ \vdots \\ \vec{p}_N^{\,\,\rm fl}\end{pmatrix}+ (\boldsymbol{T}^{-1}\boldsymbol{A})\begin{pmatrix} \vec{E}_1^{\rm b} \\ \vdots \\ \vec{E}_N^{\rm b}\end{pmatrix}.
	    \label{Eq:dipol}
\end{eqnarray}
Here we have used the auxilliary $3N\times3N$-block matrices~\cite{nteilchen} 
\begin{align}
	\boldsymbol{1}_{ij} &= \delta_{ij}\mathds{1}, \\
	\boldsymbol{T}_{ij} &= \delta_{ij}\mathds{1}-(1-\delta_{ij})k_0^2\alphamat\mathds{G}_{ij},  \\
	\boldsymbol{D}_{ij} &=\epsilon_0\mu_0\mathds{G}_{ij}
\end{align}
and
\begin{equation}
  \boldsymbol{A}_{ij} = \epsilon_0\delta_{ij}\alphamat
\end{equation}
with the Green's functions $\mathds{G}_{ij} = \mathds{G}^{\rm EE}(\vec{r}_i,\vec{r}_j)$ for the electric field due to electric sources as defined in Ref.~\cite{Eckhardt} and explicitely given in Appendix~\ref{App:Green}.

With these expressions the net mean power received by particle $i$ defined as~\cite{nteilchen}
\begin{equation}
  \begin{split}
     \langle P_{i}\rangle &= \biggl\langle\frac{d\vec{p}_i(t)}{dt} \cdot \vec{E}_{i}(t)\biggr\rangle \\
                          &= 2{\rm Im}\int_{0}^{\infty}\frac{{\rm d}\omega}{2\pi} \langle\vec{p}_i(\omega) \cdot \vec{E}_{i}^*(\omega)\rangle  
  \end{split}
  \label{panfang}
\end{equation}
can be determined by inserting the above expressions. In order to evaluate the ensemble averages $\langle \circ \rangle$ we assume local thermal
equilibrium of the background field and the particles so that we can exploit the fluctuation-dissipation theorem for the fields~\cite{nteilchen}
\begin{equation}
  \langle\vec{E}^{\rm b}_{i} \otimes \vec{E}^{\rm b^*}_{j}\rangle = 2\omega^2\mu_0\hbar\Big(n_b+\frac{1}{2}\Big)\frac{\mathds{G}_{ij}-\mathds{G}_{ji}^\dagger}{2\I} 
	\label{Eq:FDTfield}
\end{equation}
and for the fluctuating dipole moments~\cite{nteilchen}
\begin{equation}
	\langle\vec{p}_i^{fl}\otimes\vec{p}_j^{fl*}\rangle = 2\epsilon_0\hbar \delta_{ij} \Big(n_i +\frac{1}{2}\Big)\chimat
        \label{Eq:FDTdipol}
\end{equation}
Here we have introduced the mean occupation number $n_{i/b} \equiv n(T_{i/b}) = 1/(\exp(\hbar \omega / \kb T_{i/b}) - 1)$ with the reduced Planck constant $\hbar$ and Boltzmann constant $k_B$. 
With these definitions and relations we finally obtain 
\begin{equation}
\begin{split}
  \langle P_{i}\rangle &= 4\Im \int_{0}^{\infty}\frac{{\rm d}\omega}{2\pi}\hbar\omega \sum_{j=1}^{N}(n_j-n_b) \\
                       &\qquad \times {\rm Tr}\Big[\blockt_{ij}\chimat_j(\boldsymbol{DT}^{-1})_{ij}^\dagger\Big]
\end{split}
\label{pn}
\end{equation}
with the generalized suszeptibility
\begin{equation}
  \chimat_{i} = \frac{\alphamat-\alphamat^\dagger}{2\I}-k_0^2\alphamat\cdot\frac{\mathds{G}_{ii}-\mathds{G}_{ii}^\dagger}{2\I}\cdot\alphamat^\dagger
\end{equation}
of the $i$-th particle. 

In our special case of two particles which are in a plane parallel to the substrate we have $\chimat_1 = \chimat_2$ because $\mathds{G}_{11} = \mathds{G}_{22}$ due to translational symmetry. Furthermore, for $N = 2$ the mean power received by particle $1$ is 
\begin{equation}
	\langle P_{1}\rangle = 3\int_{0}^{\infty}\frac{{\rm d}\omega}{2\pi}\hbar\omega \bigl[ (n_1 - n_b) \mathcal{T}_1^a + (n_2 - n_b) \mathcal{T}_1^b \bigr]
\label{p1}
\end{equation}
introducing the transmission coefficients
\begin{align}
\mathcal{T}_1^a &= \frac{4 k_0^2}{3}{\rm ImTr}\Bigg[\mathds{D}_{121}^{-1}\chimat_1 \notag \\
                &\qquad \times \Big(\mathds{G}_{11}\mathds{D}_{121}^{-1}+\mathds{G}_{12}\mathds{D}_{212}^{-1}k_0^2\alphamat\mathds{G}_{21}\Big)^\dagger\Bigg] \\
\mathcal{T}_1^b &=\frac{4 k_0^2}{3} {\rm ImTr}\Bigg[\mathds{D}_{121}^{-1}k_0^2\alphamat\mathds{G}_{12}\chimat_2 \notag \\ 
                & \qquad \times \Big(\mathds{G}_{11}\mathds{D}_{121}^{-1}k_0^2\alphamat\mathds{G}_{12}+\mathds{G}_{12}\mathds{D}_{212}^{-1}\Big)^\dagger\Bigg]
\end{align}
with
\begin{equation}
   \mathds{D}_{iji} := \mathds{1}-k_0^4\alphamat\mathds{G}_{ij}\alphamat\mathds{G}_{ji}
\end{equation}
The expression for $\langle P_{2}\rangle$ can be obtained by interchanging $1 \leftrightarrow 2$.
In the backward case with $T_1 = T_b$ and $T_2 > T_1$, i.e.\ when particle $2$ is heated with respect to it's environment (see Fig.~\ref{Fig:Geometry}), then the power $\langle P_1 \rangle$ received by particle $1$ describes the heat flux from particle $2$ to particle $1$ which is obviously described by the transmission coefficient $\mathcal{T}_1^b$. Similarly $\mathcal{T}_2^b$ would describe the heat flux from particle $1$ to particle $2$ in the forward case that $T_2 = T_b$ and $T_1 > T_2$ (see Fig.~\ref{Fig:Geometry}). As shown explicitely in Ref.~\cite{Herz}, for reciprocal particles and substrate, i.e.\ if $\uuline{\alpha}^t = \uuline{\alpha}$ and $\mathds{G}_{ij} = \mathds{G}^t_{ji}$, we find $\mathcal{T}_1^b = \mathcal{T}_2^b$. The heat flux in forward and backward direction is the same. Now, if the particles or the environment are non-reciprocal then $\mathcal{T}_1^b \neq \mathcal{T}_2^b$ in general~\cite{Herz}. Hence, for non-reciprocal materials the heat flux in forward and backward direction are not the same. In the following we will show that when applying a magnetic field the non-reciprocal surface modes in the InSb sample will result in a large heat flux rectification.

 %
 %
\section{material properties}

For an applied magnetic field in positive $y$ direction the permittivity of InSb is given by 
\begin{equation}
   \underline{\underline{\rm \epsilon}}=\begin{pmatrix} \epsilon_1 & 0&{\rm i}\epsilon_2 \\ 0 & \epsilon_3 & 0 \\- {\rm i}\epsilon_2 & 0 & \epsilon_1 \end{pmatrix} 
\label{epsilon}
\end{equation}
with \cite{Palik}
\begin{align}
	\frac{\epsilon_1}{\epsilon_\infty} &= \left(1+\frac{\omega_{\rm L}^2-\omega_{\rm T}^2}{\omega_{\rm T}^2-\omega^2-{\rm i}\Gamma\omega}+\frac{\omega_{\rm p}^2(\omega+{\rm i}\gamma)}{\omega[\omega_{\rm c}^2-(\omega+{\rm i}\gamma)^2]} \right),
\label{eps1} \\
	\frac{\epsilon_3}{\epsilon_\infty} &= \left(1+\frac{\omega_{\rm L}^2-\omega_{\rm T}^2}{\omega_{\rm T}^2-\omega^2-{\rm i}\Gamma\omega}-\frac{\omega_{\rm p}^2}{\omega(\omega+{\rm i}\gamma)} \right)
\label{eps3}
\end{align}
and
\begin{equation}
	\frac{\epsilon_2}{\epsilon_\infty} = \frac{\omega_{\rm p}^2\omega_{\rm c}}{\omega[(\omega+{\rm i}\gamma)^2-\omega_{\rm c}^2]}
\label{eps2} 
\end{equation}
with the cyclotron frequency $\omega_c = eB/m^*$, the effective mass $m^* = 7.29\times10^{-32}$ kg, the density of the free charge carriers $n = 1.36\times10^{19}$ cm$^{-3}$~\cite{exp}, the dielectric constant for infinite frequencies $\epsilon_\infty = 15.68$, the longitudinal and transversal optical phonon frequency $\omega_{\rm L} = 3.62\times10^{13}$ rad/s and $\omega_{\rm T} = 3.39\times10^{13}$ rad/s~\cite{Palik}. With these parameters, the plasma frequency of the free carriers is $\omega_{\rm p} = \sqrt{\frac{ne^2}{m^*\epsilon_0\epsilon_\infty}} = 1.86\times10^{14}$ rad/s. Furthermore, we use the phonon damping constant $\Gamma = 5.65\times10^{11}$ rad/s an the free charge carrier damping constant $\gamma = 10^{12}$ rad/s~\cite{exp}. 

From the above expressions for the permittivity it can be seen that the permittivity tensor is diagonal, if no magnetic field is applied ($B = 0$) and therefore $\underline{\underline{\rm \epsilon}} = \underline{\underline{\rm \epsilon}}^t$. On the other hand, if $B \neq 0$ the permittivity is non-reciprocal, i.e.\ $\underline{\underline{\rm \epsilon}} \neq\underline{\underline{\rm \epsilon}}^t$, due to the Lorentz force acting on the electrons inside InSb. This will have an impact not only on the surface modes in the InSb sample but also on the localized resonances of the InSb nanoparticles.

\section{Localized magneto-optical plasmons of the nanoparticles}

In dipole approximation the polarizability of the particles is~\cite{LakhtakiaEtAl1991}
\begin{equation}
  \underline{\underline{\alpha}} = 4\pi R^3(\underline{\underline{\epsilon}}-\mathds{1})(\underline{\underline{\epsilon}}+2\mathds{1})^{-1}.
\label{alphaerg}
\end{equation}
Thus, due to the non-reciprocity of the permittivity the polarizability becomes non-reciprocal as well, if a magnetic field is applied we have $\underline{\underline{\rm \alpha}} \neq\underline{\underline{\rm \alpha}}^t$. Furthermore, as discussed in detail in Ref.~\cite{meinpaper} the three-fold degenerate localized dipolar resonances at $\omega_{m = 0,\pm 1}$ determined by the poles of $\uuline{\alpha}$ with magnetic quantum numbers $m = 0, \pm1$ split into three non-degenerated resonances where the splitting of the resonances with $m = \pm 1$ is mainly given by the cyclotron frequency $\omega_c$. As shown in Ref.~\cite{meinpaper} these resonances are connected with a clockwise (counter-clockwise) radiative heat flux for $m = -1$ ($m = +1$) as well as an angular momentum and spin which also persist in global thermal equilibrium and which are at the heart of the persistent heat current and thermal Hall effect in many-particle assemblies~\cite{zhufan,meinpaper,hall,OttEtAl2019}. As shown in Ref.~\cite{meinpaper} the mean spin for  $m = -1$ ($m = +1$) is parallel (anti-parallel) to the magnetic field resulting in a blue (red) shift so that the splitting can be understood as an analogue of the Zeeman splitting.

\section{Magneto-optical surface modes of the substrate}

The non-reciprocity introduced by the magnetic field also affects the surface modes of the InSb sample~\cite{BrionEtAl1972,wallis,chiuquinn}. To see this effect and to determine the heat transfer, we have determined the reflection matrix which has in the polarization basis of s- and p-polarization (TE and TM modes) the form
\begin{equation}
	\mathds{R} = \begin{pmatrix} r_{ss} & r_{ps} \\ r_{sp} & r_{pp} \end{pmatrix}
\end{equation}
analogously to the approach in Ref.~\cite{chen} by solving the Booker equation analytically. For $B = 0$ the diagonal elements $r_{sp}$ and $r_{ps}$ describing depolarization effects vanish. But even for $B \neq 0$ they turn out to be small compared to $r_{ss}$ and $r_{pp}$ for the parameters used in our work. Therefore, the reflection coefficients $r_{sp}$ and $r_{ps}$ can be neglected. Furthermore, to study the impact of the magnetic field on the surface modes it suffices to focus on $r_{pp}$. Considering the Voigt configuration as depicted in Fig.~\ref{Fig:Geometry} where the magnetic field is in y-direction and the surface modes travel in $\pm k_x$ direction, the reflection coefficient can be written as  
\begin{equation}
   r_{pp}=\frac{(k_0^2-k_{z}k_z^-)(k_x^2-k_0^2\epsilon_1)+k_{z}k_x(k_0^2\I\epsilon_2+k_xk_z^-)}{k_{z}k_x(k_0^2\I\epsilon_2+k_xk_z^-) - (k_0^2-k_{z}k_z^-)(k_x^2-k_0^2\epsilon_1)}\label{rppkx}
\end{equation}
introducing the wave vector components
\begin{align}
	k_{z} = \sqrt{k_0^2-k_x^2},\\
	k_{z}^- = \sqrt{k_0^2\epsilon_v-k_x^2}
\end{align}
with the Voigt permittivity 
\begin{equation}
	\epsilon_v = \frac{\epsilon_1^2 - \epsilon_2^2}{\epsilon_1}
\end{equation}
and the wavenumber in vacuum $k_0=\omega/c$ where $c$ is the vacuum light velocity. Now, the dispersion relation of the surface modes is given by the poles of $r_{pp}$. We obtain
\begin{equation}
   k_x^2-k_0^2\epsilon_1-k_{z}k_{z}^-\epsilon_1-k_{z}k_x\ri\epsilon_2=0
	\label{Eq:dispersion}
\end{equation}
which is the same expression as in Ref.~\cite{chiuquinn}. It can already be seen that this dispersion relation depends on the sign of $k_x$. Hence, surface modes propagating to positive or negative x-direction have in general different dispersion relations if $\epsilon_2 \neq 0$, i.e.\ if a magnetic field is applied. Furthermore, in the quasi-static regime $k_x^2 \gg k_0^2, k_0^2 |\epsilon_v| $ we retrieve the result~\cite{chiuquinn} $\epsilon_1 + \epsilon_2 = -1$ for $k_x > 0$ and $\epsilon_1 - \epsilon_2 = -1$ for $k_x < 0$ reflecting again the fact that the surface modes propagating in positive or negative x-direction are differently affected by the magnetic field which introduces this non-reciprocity. This non-reciprocity is more generally expressed by that fact that $r_{pp} (k_x) \neq r_{pp} (-k_x)$ when a magnetic field is applied in y-direction. As a consequence also the Green's tensor becomes non-reciprocal in this case and of course it is clear that the heat flux from particle $1$ to particle $2$ due to the coupling to the surface waves will be different from the heat flux from particle $2$ to particle $1$.

\begin{figure}
	\centering
	\includegraphics[width=0.5\textwidth]{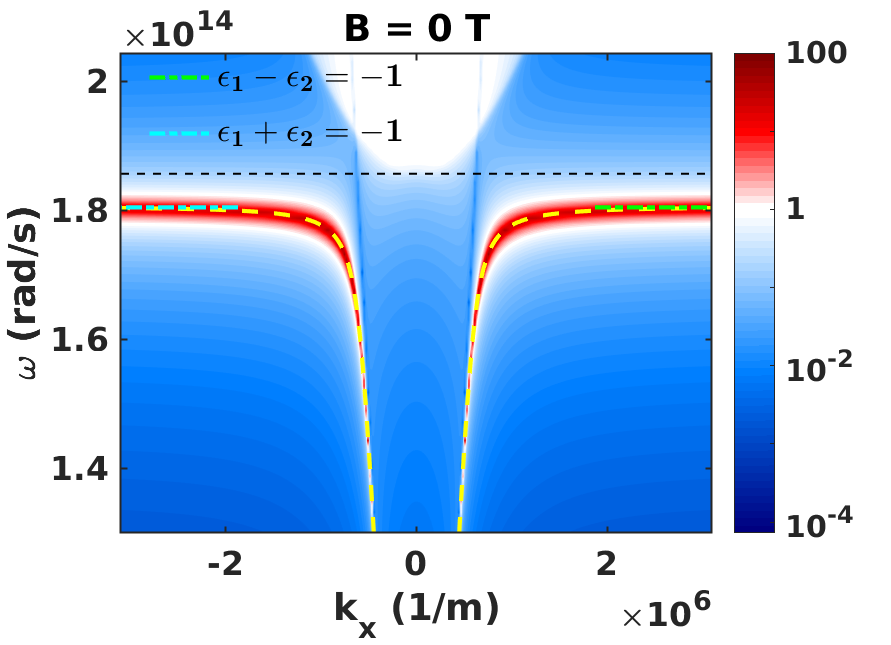}\\
	\includegraphics[width=0.5\textwidth]{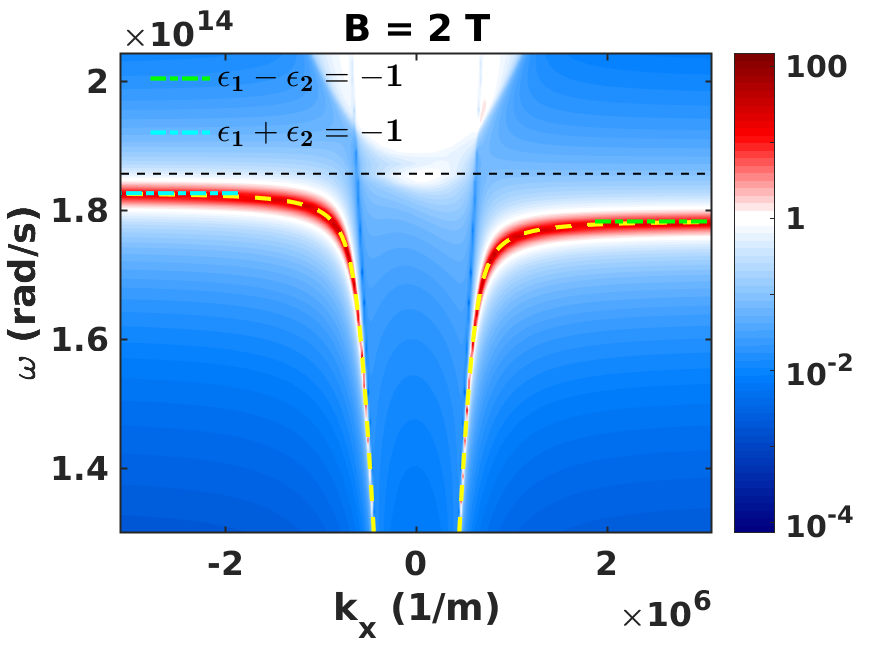}
	\caption{Parallel part of the reflection coefficient $r_{pp}$ for the Voigt configuration in $\omega$-$k_x$-plane with $k_y = 0$ and $B = 0$ (top) and $B = 2\,{\rm T}$ (bottom). For propagating waves with $k_x^2 \leq k_0^2 $ the quantitiy $1-|r_{pp}|^2$ is shown and for evanescent waves with $k_x^2 > k_0^2$ the quantity $\Im(r_{pp})$ is plotted. The yellow dashed line is the dispersion relation in Eq.~(\ref{Eq:dispersion}). The dashed green and blue curves show the quasistatic limit of the dispersion relation. The horizontal dashed line marks the plasma frequency of InSb.}
	\label{bsprpp}
\end{figure}

The non-reciprocal behaviour of the surface modes can be seen in Fig.~\ref{bsprpp} where we have plotted $1 - |r_{pp}|^2$ for the propagating waves with $k_x^2 \leq k_0^2$ and $\Im(r_{pp})$ for the evaneszent waves with $k_x^2 > k_0^2$. These quantities reflect the absorbed energy by reflection of incident propagating and evanescent waves. We have also plotted the dispersion relation of the surface modes in Eq.~(\ref{Eq:dispersion}). It can be easily seen that the symmetry of $r_{pp}$ with respect to $k_x$ is broken when a magnetic field $B \neq 0$ is applied in y direction. A splitting of the two resonances can be seen, where the surface modes travelling in positive x-direction are red-shifted whereas the surface modes travelling in negative x-direction are blue shifted~\cite{BrionEtAl1972}. As discussed in Ref.~\cite{Mechelen,Zubin2019} there is a spin-momentum locking of the surface waves. The surface waves for $k_x < 0$ have a spin in positive y-direction, i.e.\ in the direction of the magnetic field, whereas the surface waves for $k_x > 0$ have a spin in negative y-direction, i.e.\ opposite to the magnetic field. From this one can intuitively understand the red-shift of the resonance frequency for surface waves with $k_x > 0$ and the blue shift for surface waves with $k_x < 0$ which is agaim simply analoguous to the Zeeman effect.

\section{Local Density of States}

The fact, that the presence of the magnetic field introduces a non-reciprocity or assymetry for the waves propagating in positive or negative x-direction motivates to devide the expression for the local density of states (LDOS) $D(\omega,z)$ into two parts $D^\pm(\omega,z)$ belonging to exactly such waves with $k_x > 0$ and $k_x < 0$. Thus, we define $D^\pm (\omega,z)$ by starting from the well-known expression of the LDOS at a distance $z$ above a semi-infinite medium~\cite{LDOS}
\begin{equation}
	D(\omega,z) = \int_{-\infty}^{\infty}\frac{{\rm d}k_x}{2\pi}\int_{-\infty}^{\infty}\frac{{\rm d}k_y}{2\pi} f(\kappa,\omega,z)
\end{equation}
with $\kappa^2 = k_x^2 + k_y^2$ and
\begin{equation}
\begin{split}
	f(\kappa,\omega,z) &= \frac{\omega}{\pi c^2}{\rm Im}\frac{\ri}{2\sqrt{k_0^2-\kappa^2}} \\
	                    &\quad \times\Big[4+\frac{2\kappa^2}{k_0^2}\Big((r_{ss}+r_{pp})e^{2\ri\sqrt{k_0^2-\kappa^2}z}\Big)\Big].
\end{split}
\end{equation}
Strictly speaking this expression is only valid for media with $r_{sp} = r_{ps} = 0$. Since, we find for InSb with our choice of parameters that these depolarization compononts are negligible small compared to $r_{ss}$ and $r_{pp}$, we can also use the expression of the LDOS to characterize our InSb sample. Now, we define $D^\pm (\omega,z)$ by considering only the contributions for the waves travelling in positive and negative x-direction
\begin{align}
	D^+ (\omega,z) &= \int_{0}^{\infty}\frac{{\rm d}k_x}{2\pi}\int_{-\infty}^{\infty}\frac{{\rm d}k_y}{2\pi} f(\kappa,\omega,z), \label{Eq:Dplus} \\
     	D^- (\omega,z) &= \int_{-\infty}^{0}\frac{{\rm d}k_x}{2\pi}\int_{-\infty}^{\infty}\frac{{\rm d}k_y}{2\pi} f(\kappa,\omega,z). \label{Eq:Dminus}
\end{align}
In Fig.~\ref{zustandsdichte} we will use this quantity to discuss the heat flux rectification.

\section{Heat transfer mechanism}

\begin{figure}
	\centering
	\includegraphics[width=0.4\textwidth]{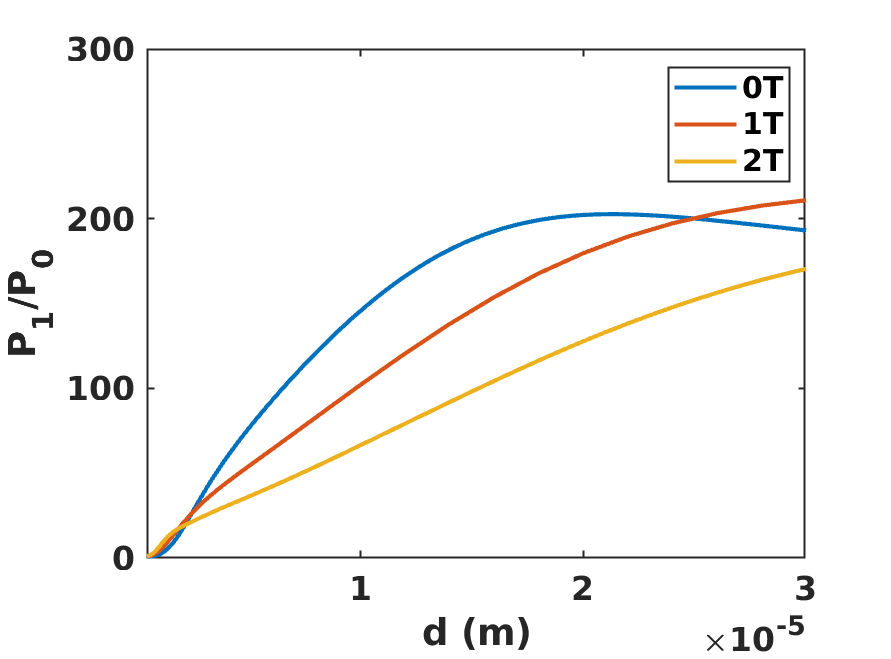}\\
	\includegraphics[width=0.4\textwidth]{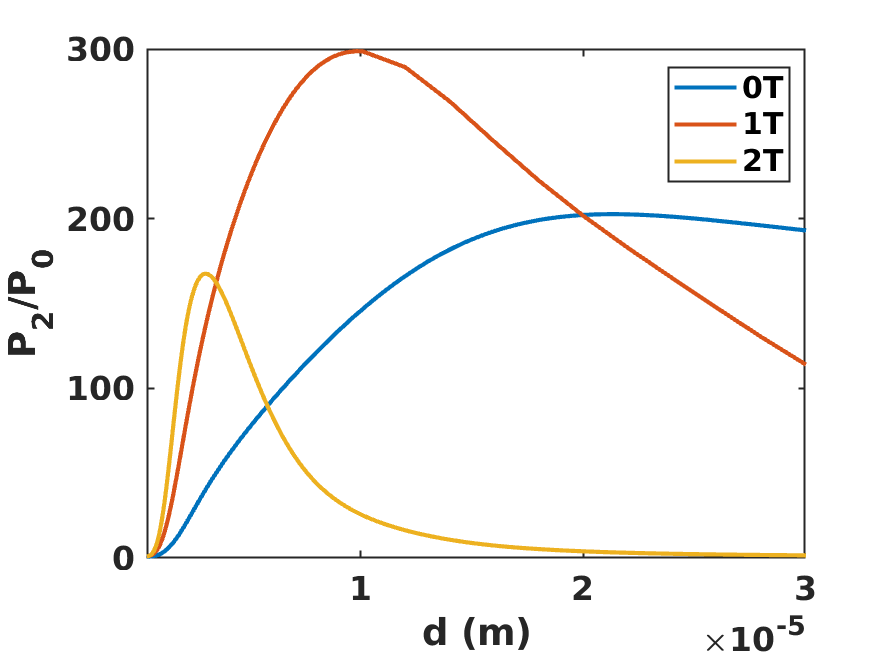}
	\caption{Net transferred power $P_1$ and $P_2$ on the colder particles in the backward and forward case (see Fig.~\ref{Fig:Geometry}) as function of interparticle distance $d$ for different strengths of the magnetic field $B = 0\,{\rm T}, 1\,{\rm T}, 2\,{\rm T}$. The distance to the substrate is $z=5R=500$nm. The transferred power is normalized to the value $P_0$ where the substrate is replaced by vacuum.}
		\label{pp0}
\end{figure}

Due to the non-reciprocal behaviour of the surface modes the heat fluxes between the two particles can become asymmetric if the heat flux is dominated by the contribution of the surface modes. To study this effect, we consider now the configuration shown in Fig.~\ref{Fig:Geometry} for the backward scenario with $T_1 = T_b = 300$K and $T_2=350$ K and the forward scenario with $T_2 = T_b = 300$K and $T_1=350$ K chosing $z = 5R = 500$nm. In Fig.~\ref{pp0} the net power $P_1 \equiv \langle P_1 \rangle$ received by particle $1$ in the backward case and the net power $P_2 \equiv \langle P_2 \rangle$ received by particle $2$ in the forward case are shown for different interparticle distances $d$. Note, that these powers are normalized to the value $P_0$ where the substrate is replaced by vacuum. It can be easily seen that $P_1 \neq P_2$ if $B \neq 0$ and that the maximum of $P_1/P_0$ at position $d_m$ moves to larger distances $d$ out of the plotted region when increasing the magnetic field amplitude, whereas the maximum of $P_2/P_0$ moves to smaller distances. 

\begin{figure}
	\centering
	\includegraphics[width=0.4\textwidth]{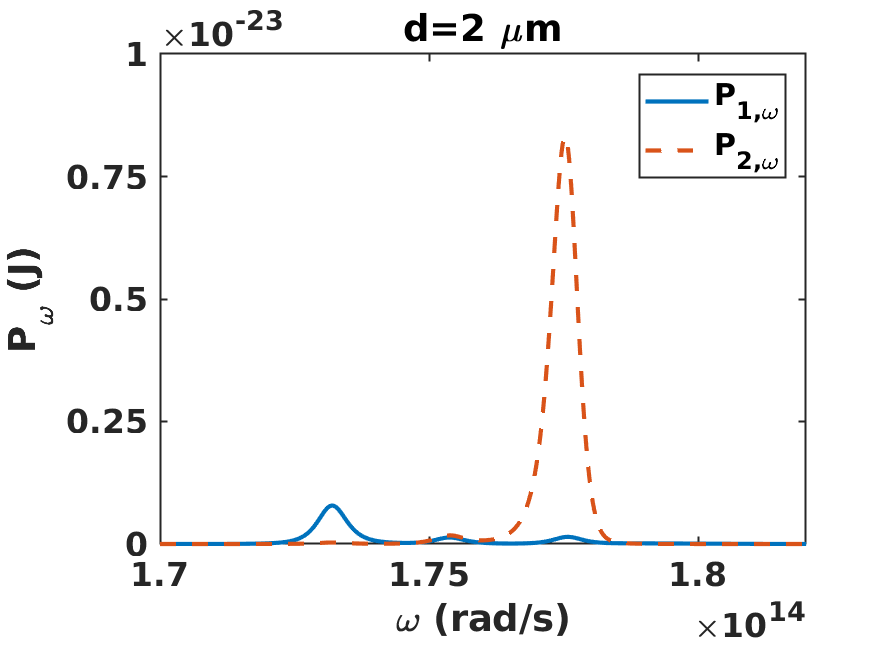}\\
	\includegraphics[width=0.4\textwidth]{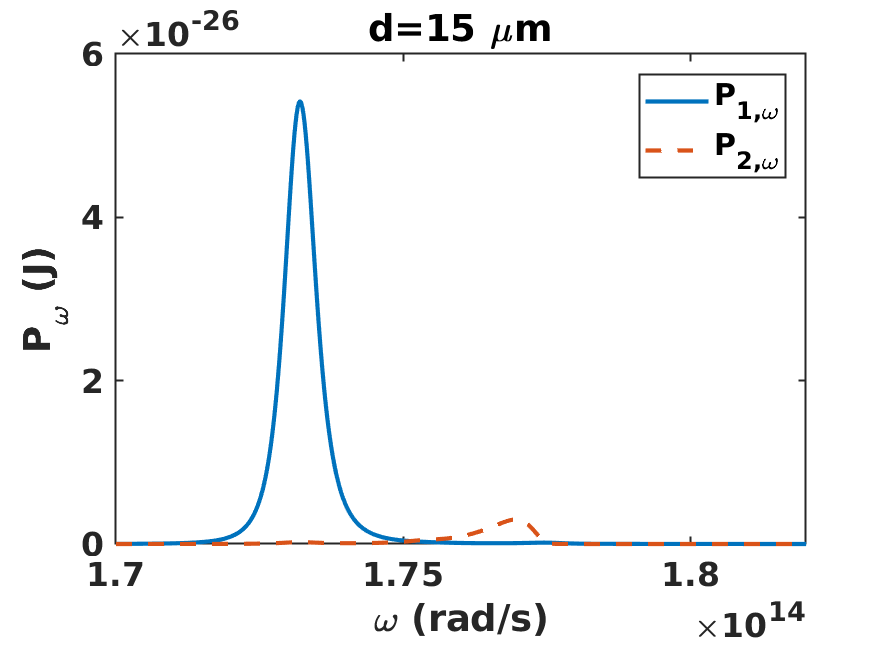}
	\caption{Spectral net power in forward and backward direction for $d = 2\,\mu{\rm m}$ and  $d = 15\,\mu{\rm m}$ and $B = 2\,{\rm T}$. The distance to the substrate is $z=5R=500$nm.}
		\label{spekma}
\end{figure}

The spectra $P_{1,\omega}$ and $P_{2,\omega}$ for the backward and forward case are shown in  Fig.~\ref{spekma} for $d=2\mue$m and  $d=15\mue$m. It is apparent that for the forward direction $P_{2,\omega}$ is dominated by the high-frequency resonance with magnetic quantum number $m = -1$ of the nanoparticles and for the backward direction $P_{1,\omega}$ is dominated by the low-frequency resonance with magnetic quantum number $m=+1$. Furthermore, it can be observed that for $d=2\mue$m we have $P_2 > P_1$ and for  $d=15\mue$m  $P_1 > P_2$. Hence, there is a clear rectification of the heat flux which changes its direction when changing from near-field to far-field interparticle distances $d$. Furthermore, since $P_2$ ($P_1$) can only be due to the coupling to the surface waves travelling in positive (negative) x-direction, this suggest that the localized particle resonance with $m = +1$ having negative spin couples preferably to the surface wave with negative spin and the particle resonance with $m = -1$ having positive spin to the surface wave with positive spin. Hence, our results suggest that there is a selection rule allowing preferred coupling between particle and surface resonances with the same spin.  

From this coupling mechanism, we can also understand the position of the maximum in $P_1/P_0$ and $P_2/P_0$. To this end, we consider now the propagation length of the surface modes, which is defined as
\begin{equation}
  \Lambda^\pm = \frac{1}{\pm 2 \Im(k_x^\pm)}
\end{equation}
where $k_x^\pm$ is the complex solution $k_x$ with $\Re(k_x) > 0$ or $\Re(k_x) < 0$ of the dispersion relation in Eq.~(\ref{Eq:dispersion}) for a given real frequency $\omega$. Note that this determines only the propagation length of surface waves with $k_y = 0$. In general, also surface waves with $k_y \neq 0$ which are included in our calculation have an impact on the heat transfer between the particles. Therefore, $\Lambda^\pm$ is only a rough estimate of the length scale of the propagation length of the surface waves contributing to the full heat transfer. 

\begin{figure}
	\centering
	\includegraphics[width=0.4\textwidth]{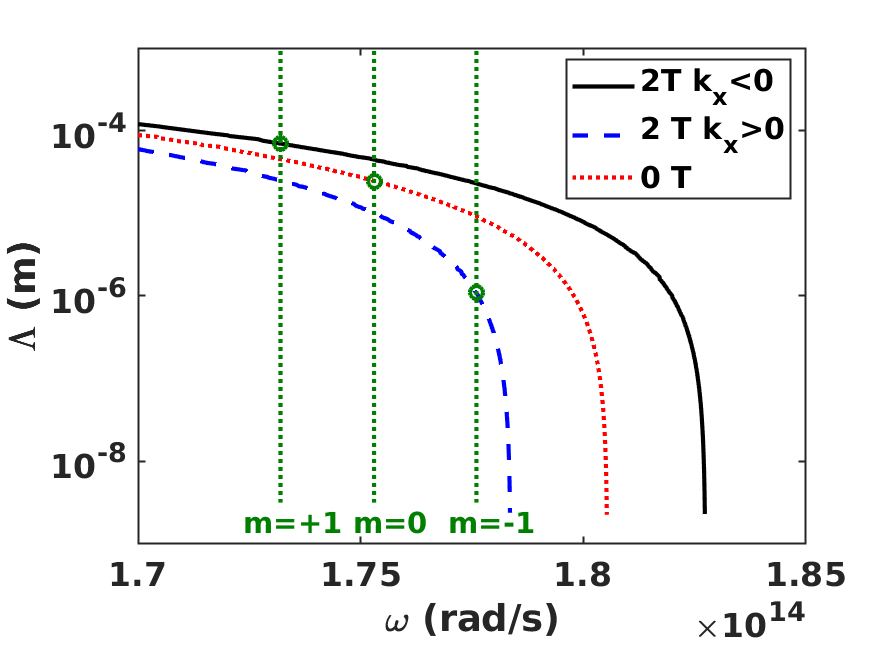}
	\caption{Propagation length $\Lambda^\pm$ of the surface modes for $B = 0\,{\rm T}$ and $2$T. The vertical lines mark the spectral position of the particle resonance with magnetic quantum numbers $m = 0,\pm1$ for $B = 2\,{\rm T}$. The circles mark the intersection of the particle resonances with the surface mode resonances.}
		\label{lambdaomega}
\end{figure}

In  Fig.~\ref{lambdaomega} we show a plot of this propagation length $\Lambda^\pm$ for the surface waves travelling in positive and negative x-direction for $B = 0\,{\rm T}$ and $2\,{\rm T}$ together with the spectral position of the three particle resonances with $m = 0, \pm 1$ for $B = 2\,{\rm T}$. Note that for $B = 0\,{\rm T}$ all resonances are at the same frequency as the $m = 0$ resonance. In Fig.~\ref{lambdaomega} it can be observed that the propagation length of the surface wave travelling in positive (negative) x-direction which couples to the $m = -1$ (m = +1) resonance has a much smaller (larger) propagation length of about $1\mu$m ($100\mu$m) for $B = 2\,{\rm T}$ than the $20\mu$m propagation length for $B = 0\,{\rm T}$. Furthermore, these values of the propagation length are in good agreement with the position $d_m$ of the maxima observed in Fig.~\ref{pp0} explaining why for small distance $P_2 > P_1$ and for large distances $P_1 > P_2$. Hence, from the coupling mechanism, the spectral shift of the particle resonances and the surface mode resonances we can understand the position of the maxima observed in $P_1/P_0$ and $P_2/P_0$ as function of the distance. A similar distance dependence has been observed for the positions of the maxima in F\"{o}rster resonance energy transfer above plasmonic surfaces~\cite{Foerster}. Furthermore, it is clear from Fig.~\ref{lambdaomega} that due to blue-shift of the $m = -1$ resonance and the red-shift of the surface modes travelling in positive x-direction, there can be no coupling anymore for large enough magnetic fields so that for large $B$ one clearly has $P_1 > P_2$ in the surface mode dominated heat transport regime.

\begin{figure}
	\centering
	\includegraphics[width=0.4\textwidth]{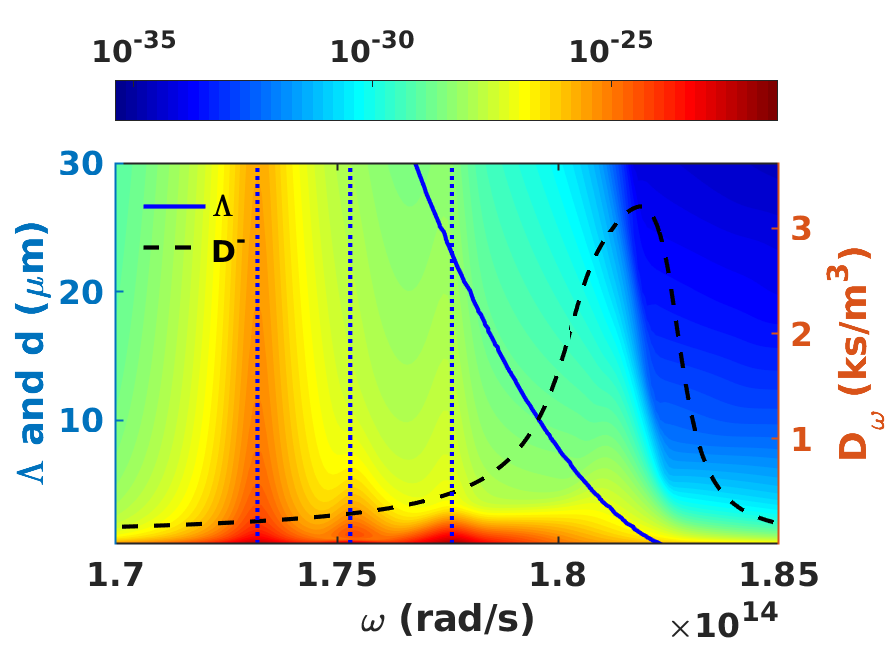}\\
	\includegraphics[width=0.4\textwidth]{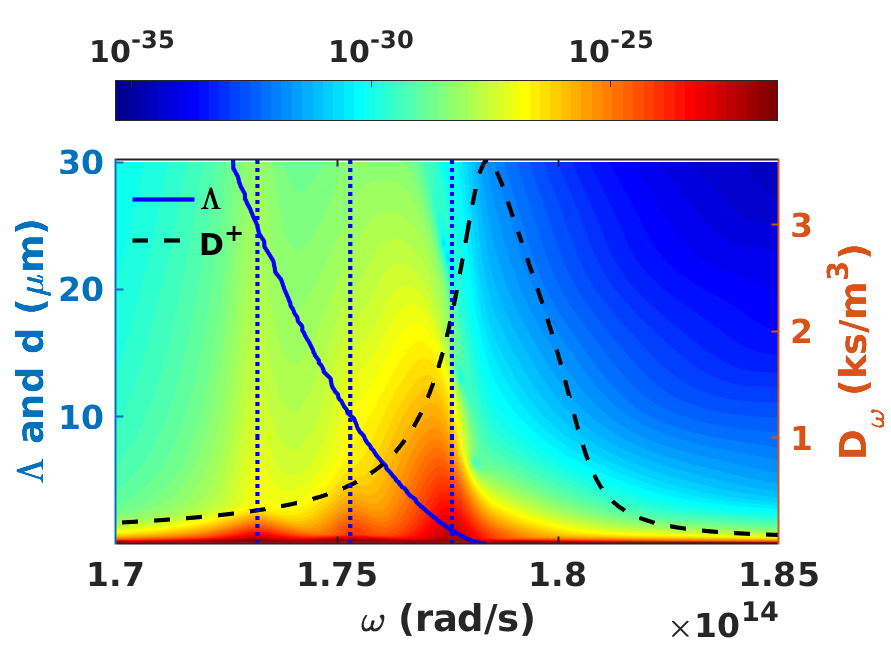}
	\caption{Spectral net power transfer $P_{1,\omega}$ (top) and $P_{2,\omega}$ (bottom) as function of interparticle distance $d$ and frequency $\omega$ for an applied magnetic field of $B =2\,{\rm T}$ and a distance of $z=5R=500$nm to the substrate. Moreover, the propagation length $\Lambda^\pm$ and the LDOS $D^\pm$ are shown. The dashed vertical lines mark the position of the particle resonance frequencies $\omega_{m = 0,\pm 1}$.}
	\label{zustandsdichte}
\end{figure}

To get a more complete pictue, in Fig.~\ref{zustandsdichte} we show the spectral power $P_{1,\omega}$ and $P_{2,\omega}$ as function of 
frequency and interparticle distance $d$ together with the position of the three particle resonances, the propagation length $\Lambda^\pm$ 
and the LDOS $D^\pm$ from Eqs.~(\ref{Eq:Dplus}) and (\ref{Eq:Dminus}). For $d \ll z$ it can be seen that all three resonances contribute
to the heat transfer due to the fact that the heat flux is mainly directly transfered between both particles. For larger $d$ the
heat transfer between both particles is more and more dominated by the coupling to the surface modes where the $m = +1$ resonance
couples to the long range surface mode travelling to negative x-direction and the $m = -1$ resonance couples to the short range
surface mode travelling to the positive x-direction. For distances much larger than $\Lambda^\pm$ the surface mode contribution
vanishes. That $P_2 > P_1$ for small distances $d$ can now also be understood by the fact that the LDOS $D^+$ is larger at 
$\omega_{m = -1}$ than $D^-$ at $\omega_{m = +1}$. For large distances $d > \Lambda^+$ only the long range surface modes can contribute
and therefore $P_1 > P_2$.

\section{Heat flux Rectification}

\begin{figure}
	\centering
	\includegraphics[width=0.4\textwidth]{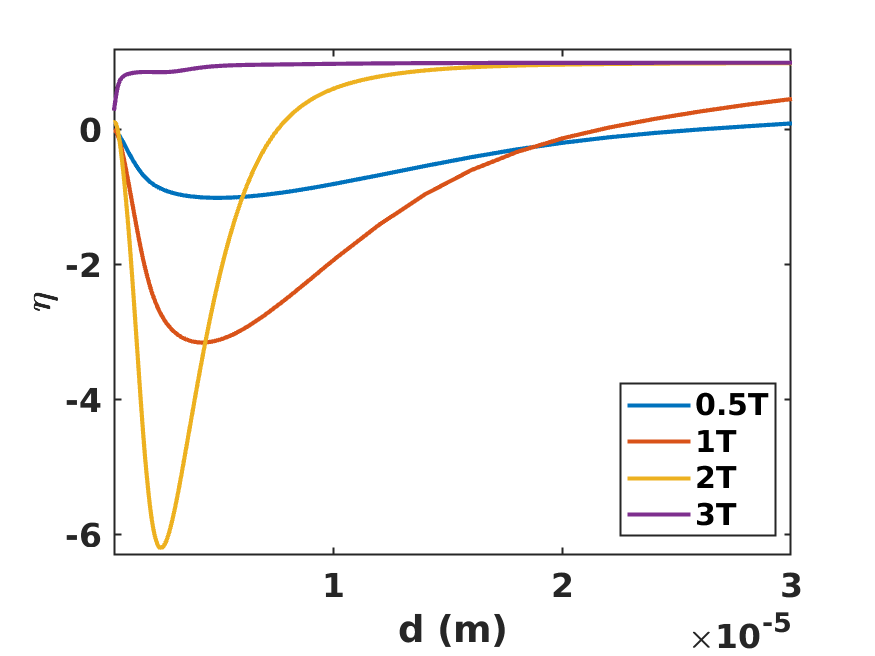}\\
	\caption{Rectification coefficient $\eta$ over the distance $d$ between the particles for different strenghts of the magnetic field $B = 0.5, 1,  2,  3\,{\rm T}$. Again we choose $z=5R=500$nm.}
	\label{etaB}
\end{figure}

To quantify the heat flux rectification we define the rectification coefficient as~\cite{paper_diode} 
\begin{eqnarray}
	\eta=\frac{P_1-P_2}{P_1}.
\end{eqnarray}
In Fig. \ref{etaB} we plot the rectification coefficient as function of $d$ for different magnetic field strengths. It can be observed that for field amplitudes smaller than 3T the rectification coefficient is negative, because $P_2 > P_1$. For large field amplitudes like $B = 3$T the rectification coefficient is purely positive, because $P_1 > P_2$. As discussed before, this is due to the fact that with increasing field strength the resonance $\omega_{m=-1}$ is blue shifted and the surface mode resonance is red-shifted leading to a decreasing propagation length $\Lambda^+$. This behaviour can here be observed in the shift of the position where $\eta = 0$ to smaller distances $d$ when the field amplitude is increased. For $B = 3$T the particle resonance  $\omega_{m=-1}$ can simply not couple to a surface mode propagating to positive x-direction anymore, because the red-shift of the particle resonance and the blue shift of the surface mode resonance are too large. Hence, for 3T we find $P_1 > P_2$ for all distances. The curves for 2T and 3T converge for $d \gg \Lambda^+$ to a rectification coefficient which is close to 1 ($\eta = 0.996$) which means that $P_1 \gg P_2$ which is a clear diode effect. On the other hand, for relatively weak fields and small distances $d$ we have a ``minimal'' $\eta$ of about $-6$ which simply means that $P_2 \approx 7 P_1$. If we would in this case define the rectification coefficient as $\tilde{\eta} = (P_2 - P_1)/P_2$ we would obtain $\tilde{\eta} \approx 0.86$. Hence, also in this case we have a large rectification, but in the other direction. 

\section{Mean Poynting vector}

To have a deeper understanding of the heat flow and the spin-spin coupling we determine now the mean poynting vector $\langle \vec{S} \rangle$ due to the thermal radiation of the two particles in a given environment. As before we provide the general expression for an arbitrary number $N$ of nanoparticles, first, and then invoke the special case $N = 2$. With the electric field 
\begin{equation}
  \vec{E}(\mathbf{r}) = \omega^2\mu_0 \sum_{i = 0}^N \mathds{G}^{\rm EE}(\mathbf{r},\mathbf{r}_i)  \vec{p}_i + \vec{E}^b(\mathbf{r})
\end{equation}
produced by the thermal background radiation and the thermal dipole moments of the nanoparticles in Eq.~(\ref{Eq:dipol}) we can determine directly the magnetic field by Faraday's law $\vec{H} = \nabla\times\vec{E}/(\ri \omega \mu_0)$. Then the mean Poynting vector $\langle\vec{S}_{\omega}\rangle = 2{\rm Re}\langle\vec{E}_\omega\times\vec{H}_\omega\rangle$ can be straight-forwardly determined by using the fluctuation-dissipation theorem of the fluctuational dipole moments in Eq.~(\ref{Eq:FDTdipol}) and of the fields in Eq.~(\ref{Eq:FDTfield}) and~\cite{Agarwal} 
\begin{equation}
   \langle\vec{E}^{\rm b}_i \otimes \vec{H}^{\rm b}_j \rangle = 2\omega^2\mu_0\hbar\Big(n_b+\frac{1}{2}\Big)\frac{\mathds{G}^{\rm EH}_{ij}-\mathds{G}^{\rm HE^\dagger}_{ji}}{2\ri}
  \label{Eq:FDTfieldb}
\end{equation}
We obtain 
\begin{equation}
\begin{split}
	\langle S_{\omega,\alpha}\rangle &=4\hbar\omega^2\mu_0k_0^2 \sum_{\beta,\gamma = x,y,z} \epsilon_{\alpha\beta\gamma} {\rm Re}\Big[ \\
	&\quad \sum_{i,j,k=1}^{N}(n_j-n_b)\mathds{G}^{\rm EE}_{0i}\blockt_{ij}\chimat_{j}\bigl(\mathds{G}^{\rm HE}_{0k}\blockt_{kl}\bigr)^\dagger \\
	&\quad  +\frac{n_b}{2 \ri} \sum_{i,j=1}^{N} \biggl(\mathds{G}^{\rm EE}_{0i}\blockt_{ij}\alphamat\mathds{G}^{\rm EH}_{j0} 
	-\big(\mathds{G}^{\rm HE}_{0i}\blockt_{ij}\alphamat\mathds{G}^{\rm EE}_{j0}\big)^\dagger \biggr) \\
&\quad  +\frac{n_b}{k_0^2}\Big(\frac{\mathds{G}^{\rm EH}_{00}-\mathds{G}^{\rm HE^\dagger}_{00}}{2\ri}\Big) \Big]_{\beta\gamma}.
\end{split}
\label{slang}
\end{equation}
where $\epsilon_{\alpha\beta\gamma}$ is the Levi-Civita tensor, $\mathds{G}^{\rm EE}_{0i} = \mathds{G}^{\rm EE} (\mathbf{r},\mathbf{r}_i)$ and $\mathds{G}^{\rm HE}_{0k} = \mathds{G}^{\rm HE}_{0k}  (\mathbf{r},\mathbf{r}_k)$, etc.\ are the electric and magnetic Green Functions of electric and magnetic sources as defined in Ref.~\cite{Eckhardt}. The first term determines the heat transfer between the particles and their environment, the third term describes only the background contribution which is for reciprocal backgrounds zero. This term describes the heat flow in the case that there are no nanoparticles. As shown in Ref.~\cite{Silveirinha} for a non-reciprocal medium this contribution persist even in global equilibrium. This persistent heat flux at an interface has also been observed for a single or several nanoparticles~\cite{meinpaper,zhufan}. Here, in particular the non-reciprocal surface modes in the InSb substrate will produce a persistent heat flux in positive or negative x-direction described by this third term. Since this term fulfills $\nabla \cdot \langle \vec{S} \rangle = 0$ it does not contribute to heat transfer between the particles~\cite{Silveirinha}. Finally, the second term is an interference term between the background field and the particles. It describes the change of heat flow of the background field due to the presence of the particles. 

\begin{figure*}
	\centering
	\includegraphics[width=0.435\textwidth]{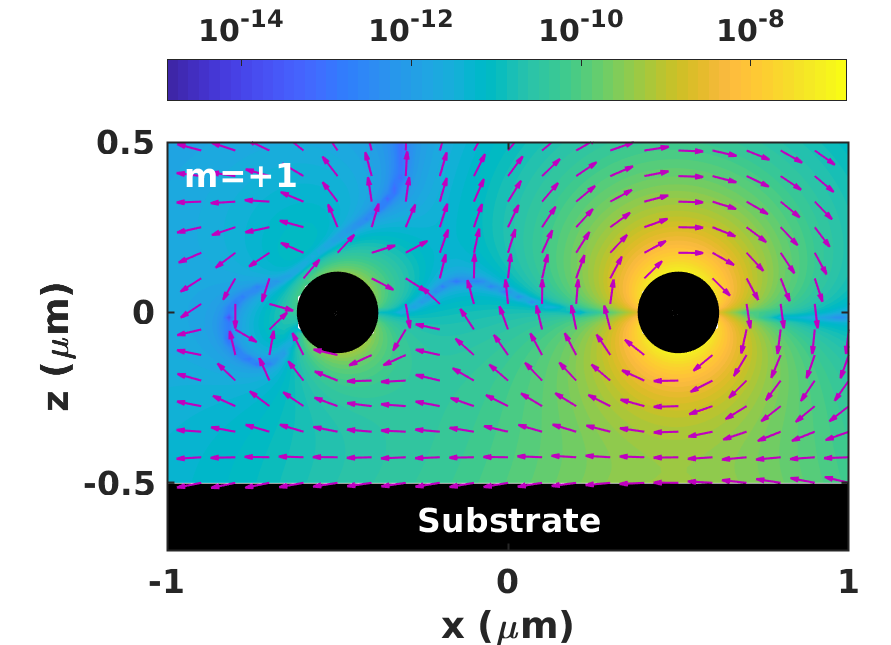}
	\includegraphics[width=0.435\textwidth]{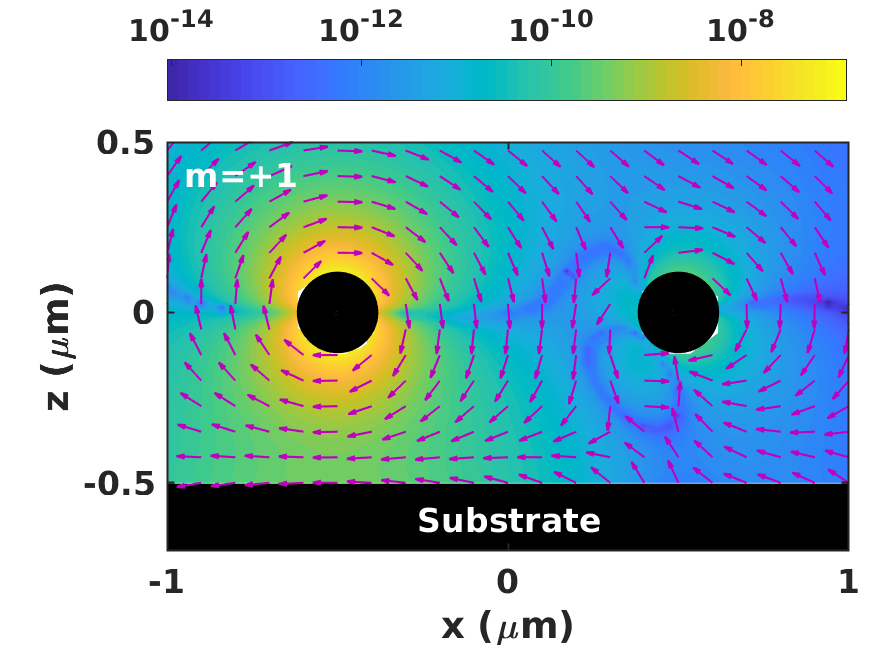}\\
	\includegraphics[width=0.435\textwidth]{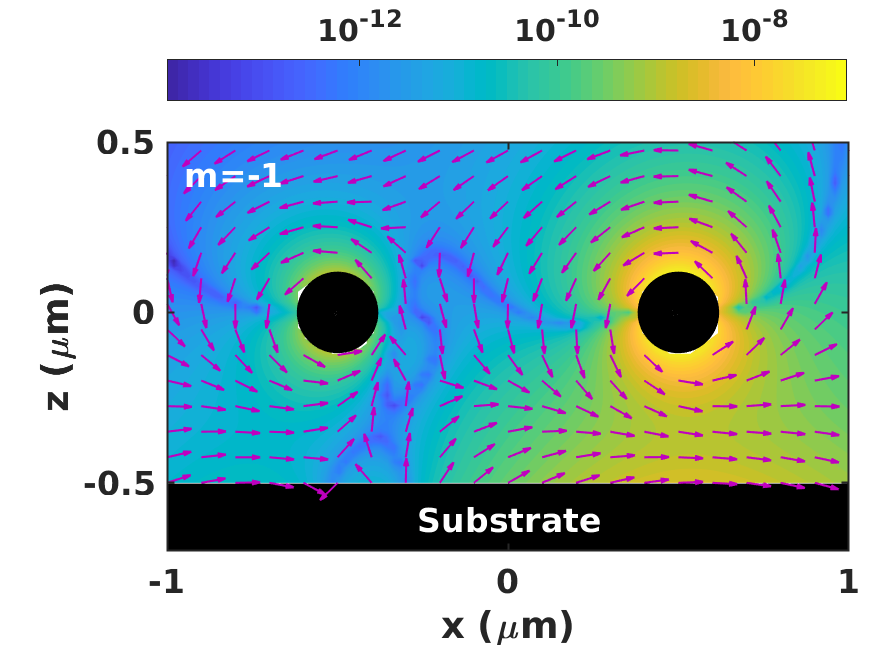}
	\includegraphics[width=0.435\textwidth]{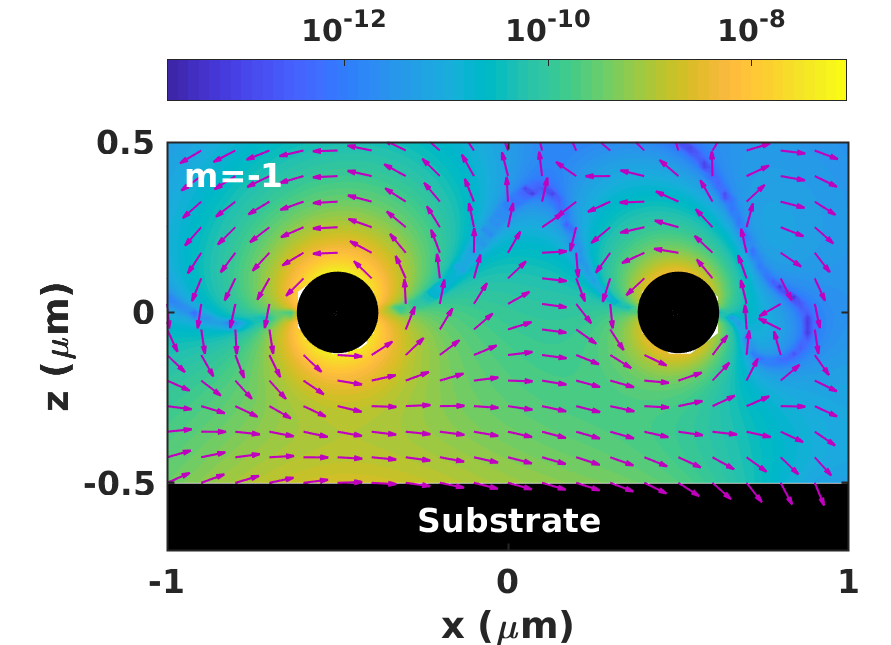}
	\caption{Spectral Poynting vector $|\langle \vec{S}_{\omega}^{\rm tr} \rangle|$ in (W/m$^2$) (colour scale) and its normalized direction (arrows) for the two InSb nano-particles (black) with a distance of $z = 5R=500$ nm to the InSb substrate and interparticle distance $d = 1\mu$m for an applied magnetic field of 2T. On the left side we show the backward case with $T_1=T_b=300$ K and $T_2=350$ K and on the right side the forward case with $T_2=T_b=300$ K and $T_1=350$ K.}
	\label{poynting_oberflaeche2Tres}
\end{figure*}

Since, the second and third term persists in global equilibrium they describe the persistent heat flux of the interface~\cite{Silveirinha} and the nanoparticles~\cite{meinpaper,zhufan}, but they do not describe any heat transfer between the nanoparticles. Therefore the heat transfer between the particles is fully determined by the first term of the Poynting vector
\begin{equation}
\begin{split}
	\langle S_{\omega,\alpha}^{\rm tr} \rangle &= 4\hbar\omega^2\mu_0k_0^2 \sum_{\beta,\gamma = x,y,z}\epsilon_{\alpha\beta\gamma}{\rm Re}\Big[ \\
	                                           &\qquad \sum_{ijk=1}^{N}(n_j-n_b) \mathds{G}^{\rm EE}_{0i}\blockt_{ij} \chi_{j}(\mathds{G}^{\rm HE}_{0k}\blockt_{kl})^\dagger\Big]_{\beta\gamma}. 
\end{split}
\label{poyntingsurf} 
\end{equation}
We have checked that for $N = 2$ the integration of the normal component of this Poynting vector on the surface of the nanoparticles $1$ and $2$ in the backward and forward scenario gives either $P_1$ or $P_2$. In Fig.~\ref{poynting_oberflaeche2Tres} $\langle \vec{S}_{\omega}^{\rm tr} \rangle$ is shown for the different resonance frequencies $\omega_{m = \pm 1}$ of the nanoparticles. Again as discussed for a single nanoparticle in Ref.~\cite{meinpaper} the mean Poynting vector is circulating around the nanoparticles clockwise (conter-clockwise) for $m = -1$ ($m = +1$). Moreover, the influence of the substrate can be clearly identified. It can be easily seen that in the forward case the net heat transfer for the $m=+1$-mode is much better then for the $m=-1$-mode and for the backward case it is the other way round. From the clockwise circularity of the $m = -1$ particle resonance it is clear that it couples preferably to surface waves with $k_x < 0$, whereas the counter-clockwise circularity of the $m = +1$ particle resonance clearly suggests a preferred coupling to surface waves with $k_x > 0$. Hence, the mean Poynting vector visualises nicely the spin-spin coupling mechanism of the circular mode in the nanoparticle and the surface modes. Furthermore there seems to be a slight heat flow from the surface towards the nanoparticles for the $m = -1$ resonance in the backward case similar to the heat pumping found in Ref.~\cite{PBA2020}

\section{Particle heating --- Where is the heat going?}

We have seen that the non-reciprocal surface modes have a strong impact in the heat flux rectification. Furthermore, it could be observed that the heat flux between the nanoparticles is strongly enhanced by the presence of the surface modes as already studied in reciprocal media~\cite{Saaskilathi2014,Asheichyk2017,DongEtAl2018,paper_2sic,ZhangEtAl2019,HeEtAl2019,MessinaEtAl2012,MessinaEtAl2016,ZhangEtAl2019b}. The evident question is the following: Is this enhanced heat flux resulting in an enhanced heating effect? Of course, when bringing the two nano-particles closer to a substrate they have a new heat flux channel namely the surface modes to exchange heat. On the other hand, they will also radiate more heat into the substrate and one can expect that this effect will be dominant~\cite{Tschikin}. Therefore to investigate the diode effect by the non-reciprocal surface waves in more detail we numerically calculate the particle temperatures in the NESS. We will assume that the temperatures of the hot particle and the background will be fixed to $310$K and $300$K in the forward and backward case. The temperature of the cold particle (particle $2$ in the forward case and particle $1$ in the backward case) which is at $t=0$s set to the same value as the background temperature, i.e.\ $300$K, will be determined by solving the energy balance equation~\cite{Tschikin}
\begin{equation}
  \rho C V \frac{dT_k}{dt} = \langle P_{k}(t,T_1, T_2,T_b)\rangle, 
\label{dgltemp}
\end{equation} 
for $k = 1,2$ with the heat capacity $C=200$ J/kg K, a mass density $\rho=5775$ kg/m$^3$ and volume $V$ of InSb nanoparticles~\cite{Piesbergen1963}. 

\begin{figure}[H]
	\centering
	\includegraphics[width=0.35\textwidth]{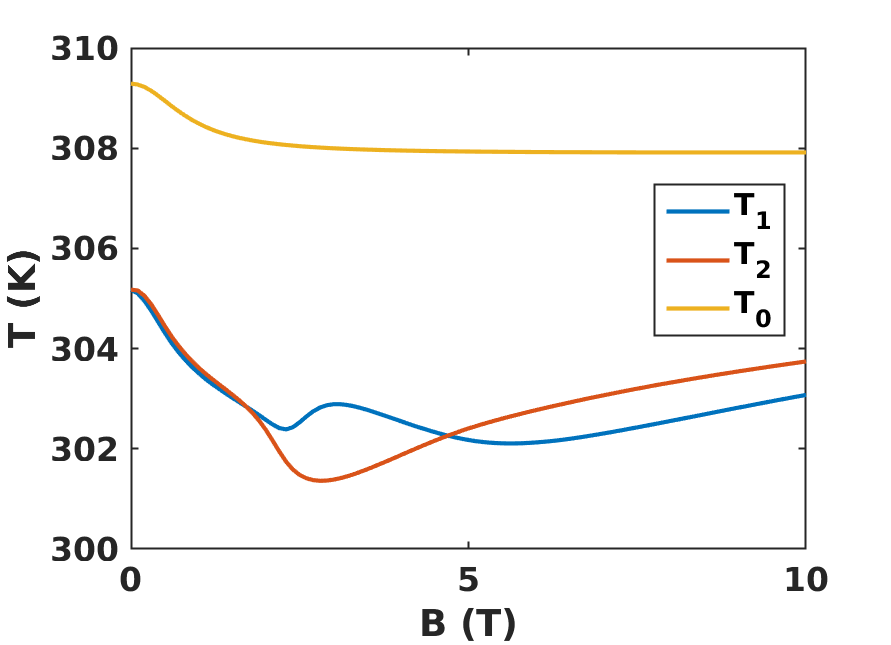}
	\caption{NESS temperatures of the coulder particle in the forward and backward direction without substrate $T_0 = T_1 = T_2$ and with substrate $T_1 \neq T_2$. We choose  $d=z=5R=500$ nm.} 
	\label{NESS}
\end{figure}

In Fig. \ref{NESS} we show the temperature of the coulder particles in the NESS for the forward and backward case with and without substrate. First, it can be observed that $T_1 = T_2 \equiv T_0$ in forward and backward direction without substrate. Here, the particle temperature of the colder particle is slightly dropping from $309$K to $308$K when the magnetic field amplitude is increased. This temperature drop of 1K (10\% of the applied temperature difference $\Delta T = 10$K) is the giant magnetic resistance effect~\cite{Latella2017, Cuevas}. When the particles are brought in the close vicinity of the substrate at $z = 500$nm, then the temperature of the colder particle for $B = 0$T drops from $309$K to about $305$K. Hence, a substantial part of the heat emitted by the warmer particle is dissipated in the substrate. Now, when turning on the magnetic field and increasing its amplitude, the temperatures first drop much faster than without surface to about $302$K until $B = 2$T and then for larger amplitudes they rise again up to $303$K/$304$K. Hence the surface enhances the giant magnetic resistance effect which is for  $B = 2$T about 30\% of the applied temperature difference, i.e.\ a temperature drop of $3$K. 

\begin{figure}[H]
	\centering
	\includegraphics[width=0.35\textwidth]{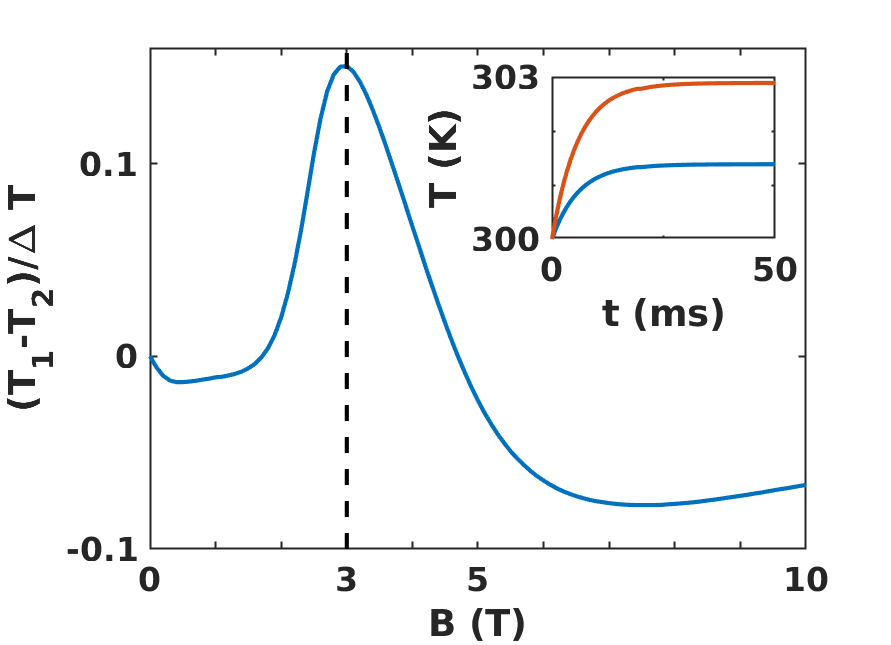}
	\caption{The temperature difference $(T_1-T_2)/\Delta T$ for the NESS temperature of the colder particle in backward and forward direction normalized to the initially applied temperature difference of $\Delta T = 10$K. The inset shows the temperature evolution of $T_2$ (blue line) and $T_1$ (red line) in forward and backward case, resp., from the initial state of $300$K to the NESS for the maximum rectification at 3T. The distance between the particles and between the particles and the substrate are $d=z=5R=500$ nm.}

	\label{relax}
\end{figure}

It can also be seen, that with substrate $T_1 \neq T_2$ in general. In Fig.~\ref{relax} we show the temperature difference $T_1 - T_2$ of the backward and forward NESS temperatures normalized to $\Delta T = 10$K. A maximum rectification of about 15\% can be observed for relatively large magnetic field amplitudes of $B = 3$T. For weak fields the effect is only -1.3\%. This is in agreement with the Hall effect, which is also relatively weak for InSb~\cite{OttEtAl2019}. Hence, a clearly measurable non-reciprocity in the heating can be observed, but due to the fact that most of the heat is going into the substrate, this effect is rather small, but we have not made any optimization procedure. We find, that the non-reciprocal heating effect cannot be simply enhanced by for example increasing $d$ or decreasing $z$. By decreasing $z$ the surface mode coupling will be stronger, but also the amount of heat going into the substrate. Also decreasing/increasing $d$ is not a priori a good option. When decreasing $d$ then the non-reciprocity vanishes due to the fact that for $d \ll z$ the coupling via the surface vanishes. Increasing $d$ on the other hand, will result in large  $P_1/P_0$ and $P_2/P_0$ as shown in Fig.~\ref{pp0} for $d =1\,\mu{\rm m}$, for instance, but the absolute value of the heat flux drops enormeously with $d$ so that the heating of the colder particles becomes unefficient. Detailed parameter studies which are out of the scope of our work of the impact of the distance $d$ and $z$ particles sizes $R$ and material properties are needed to find optimal materials and configurations to have a strong rectification effect.

\section{conclusion}

In summary, we have made a detailed discussion and investigation of the diode working principle of the non-reciprocal near-field diode in Ref.~\cite{paper_diode}. We showed that the rectficiation effect occurs due to the spin-sensitive coupling of the particle resonances and the surface modes. The transferred heat flux is maximal whenthe distance between the particle is on the order of the propagation length of the surface modes which is different for the surface modes travelling in positive or negative x direction. Moreover, the strength and direction of the effect is highly dependent on the magnetic field, the distance between the particles and the substrate as well as the local density of states. Our investigation of the mean Poynting vector showed that the spin-sensitive coupling can be understood by the circularity of the particle resonances and the directionality of the surface mode resonances. When the spin of both resonances is the same, these directionalities match explaining the spin-spin coupling. In addition, we find large rectifications of the heat flux with our choice of parameters. Nonetheless, the effective assymmetry in the heating of the nanoparticles in non-equilibrium steady state is for relatively large fields maximally 15\% of the initially applied temperature difference between the warm and could nano-particle. We believe that the search for optimal parameters and proper materials can lead to a highly increased rectification effect which would make the here discussed concept interesting for future applications. 

\appendix

\section{Dyadic Green's functions} \label{App:Green}

The Greens function of the electric field generated by the electric source currents is a sum of the vacuum $\mathds{G}^0_{ij}$ and the scattered contribution $\mathds{G}^{sc}_{ij}$~\cite{paper_2sic}:
\begin{equation}
  \mathds{G}_{ij}=\mathds{G}^{0}_{ij}+\mathds{G}^{sc}_{ij}.
\end{equation}
Here, we use the indices $i$ and $j$ to calculate the Greens function at position $\vec{r}_i=(x_i,y_i,z_i)^T$ generated by a dipole at position $\vec{r}_j=(x_j,y_j,z_j)^T$. 

For the Greens function in vacuum we use \cite{meinpaper}
\begin{equation}
  \mathds{G}^0_{ij} = \frac{e^{{\rm i}k_0 d}}{4\pi d} \left[a\mathds{1}+b\vec{e}_{d}\otimes\vec{e}_{d}\right] 
\label{greensche_funktion}
\end{equation}
with $d =|\vec{r}_i-\vec{r}_j|$ and
\begin{align}
  a &= 1+\frac{{\rm i}k_0 d-1}{k_0^2 d^2} \label{aa} \\
  b &= \frac{3-3{\rm i}k_0 d-k_0^2 d^2}{k_0^2 d^2} \label{bb}.
\end{align}

The scattered contribution due to the presence of the flat surface is given by
\begin{equation}
  \mathds{G}^{sc}_{ij} = \int_{-\infty}^{\infty}\int_{-\infty}^{\infty}\frac{{\rm d}k_x{\rm d}k_y}{(2\pi)^2}e^{{\rm i}(\vec{x}_i-\vec{x_j})\cdot\vec{\kappa}}\tilde{\mathds{G}}^{sc}_{ij}(k_x,k_y) \label{gsca}
\end{equation}
with
\begin{equation}
  \tilde{\mathds{G}}^{sc}_{ij}(k_x,k_y) = \frac{2{\rm i}e^{{\rm i}k_{i,z}(z_i+ z_j)}}{2k_{i,z}} \sum_{k,l = s,p} r_{kl}\vec{a}_k^+\otimes\vec{a}_l^-  
\label{gsci}
\end{equation}
using the polarization vectors for $s$ and $p$ polarization
\begin{equation}
  \vec{a}_s^\pm=\frac{1}{\kappa}
  \begin{pmatrix}
  k_x \\ k_y \\ 0
  \end{pmatrix}
\end{equation}
and
\begin{equation}
  \vec{a}_p^\pm=\frac{1}{\kappa k_0}\begin{pmatrix} \mp k_{i,z}k_x \\\mp k_{i,z}k_y \\ \kappa^2
  \end{pmatrix}
\end{equation}
with $\vec{x}_i=(x_i,y_i)^T$, $\vec{\kappa}=(k_x,k_y)^T$, and $k_{i,z}=\sqrt{k_0^2-\kappa^2}$. The other Green's tensors can be 
easily calculated from this expression~\cite{Eckhardt}.

\section{Impact of phonon contribution}

The effects discussed in this work highly depend on the material properties. We have chosen throughout the manuscript a parameter set for InSb with a clear dominating electric permittivity. However, for other parameter sets the phononic part may play an important role. Actually, the single surface mode band as seen in the reflection coefficients in Fig.~\ref{bsprpp} can split into several bands as found in our previous work on the diode effect in Ref.~\cite{paper_diode}, for instance. The discussion given in this work can still be applied to this case, but of course the whole picture becomes more complex. 

\begin{figure}[H]
	\centering
	\includegraphics[width=0.4\textwidth]{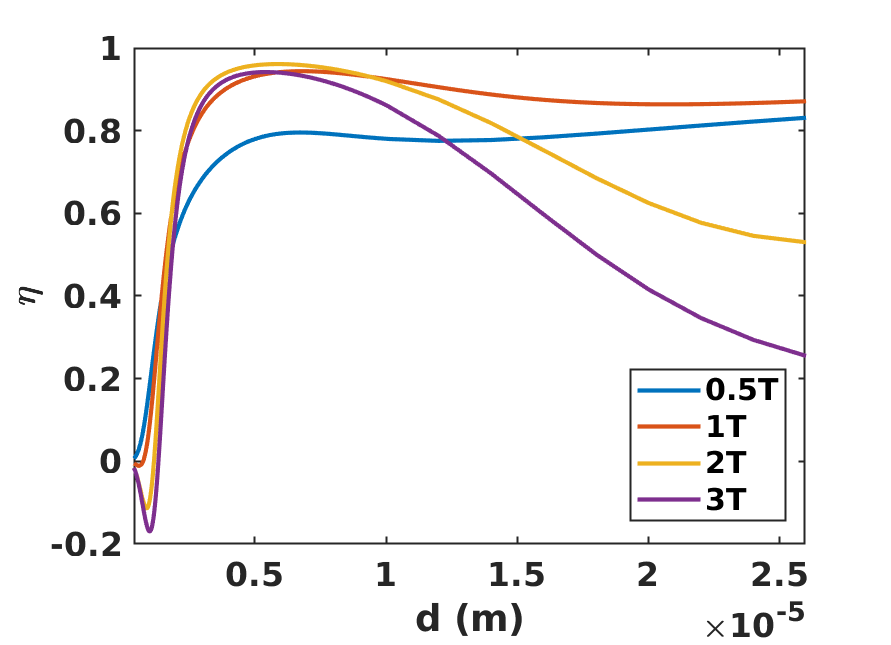}
	\caption{Rectification coefficient $\eta$ for a parameter set of InSb as used in Ref.~\cite{paper_diode} taken from~\cite{Palik} as function of interparticle distance $d$ for different strenghts of the magnetic field and $z=5R=500$nm.}
	\label{etaBcu}
\end{figure}

In order to contrast the impact of the phonons on the rectification coefficient we take now another set of parameters from Ref.~\cite{Palik} with effective mass $m^* = 1.99\times10^{-32}$ kg, density of the free charge carriers $n = 1.07\times10^{17}$ cm$^{-3}$, high frequency dielectric constant $\epsilon_\infty = 15.7$, longitudinal and transversal optical phonon frequency $\omega_{\rm L} = 3.62\times 10^{13}$ rad/s and $\omega_{\rm T} = 3.39\times10^{13}$ rad/s. Furthermore, the plasma frequency of the free carriers is $\omega_{\rm p} = \sqrt{\frac{ne^2}{m^*\epsilon_0\epsilon_\infty}} = 3.15\times10^{13}$ rad/s. We use further the phonon damping constant $\Gamma = 5.65\times10^{11}$ rad/s and the free charge carrier damping constant $\gamma = 3,39\times10^{12}$ rad/s. For this set of parameters which we have used in Ref.~\cite{paper_diode} the rectification coefficient is shown in Fig.~\ref{etaBcu}. In comparison to Fig. \ref{etaB} it can be seen that in this case the field dependence is much different. In particular, in most cases $P_1 > P_2$. As could be a priori expected, the rectification effect strongly depends on the doping level of InSb and in particular the directionality. Hence, the rectification effect can be efficiently tailored by changing the doping level. Nonetheless, it should be kept in mind that for nano-particles also size effects might play a role. For example, for nano-particles of radius $R = 100$nm the number of electrons for $n = 1.07\times10^{17}$ cm$^{-3}$ is only $450$, whereas for $n = 1.36\times10^{19}$ cm$^{-3}$ it is $56984$, i.e.\ relatively high. Hence, the optical response of the nanoparticles with a comparably low free charge carrier density like $n = 1.07\times10^{17}$ cm$^{-3}$ might be quite different from the bulk response.

\acknowledgments
\noindent

A.\ O.\ and S.-A.\ B.\ thank P. Ben-Abdallah, R. Messina, and A. Kittel for helpfull discussion and comments. S.-A.\ B.\ acknowledges support from Heisenberg Programme of the Deutsche Forschungsgemeinschaft (DFG, German Research Foundation) under the project No. 404073166.

\end{document}